\documentclass[11pt]{article}
\usepackage{moriond,epsfig}

\def\Journal#1#2#3#4{{#1} {\bf #2}, #3 (#4)}

\def\NPB{{\em Nucl. Phys.} B}
\def\PLB{{\em Phys. Lett.}  B}

\def\CPC{{\em Comp. Phys. Commun.}}

\def\be{\begin{equation}}
\def\ee{\end{equation}}
\def\bea{\begin{eqnarray}}
\def\eea{\end{eqnarray}}

\begin{document}
\rightline{\small \tt FNT/T-2000/16}
\vspace*{4cm}
\title{NON ABELIAN GAUGE COUPLINGS IN \\
FOUR FERMION PROCESSES AT LEP~\footnote{Presented by Fulvio Piccinini 
at IVth Rencontres du Vietnam: International Conference on Physics at 
Extreme Energies (Particle Physics and Astrophysics), Hanoi, Vietnam, 19-25 
July 2000.}
}

\author{G. MONTAGNA$^1$, M. MORETTI$^2$, O. NICROSINI$^1$, 
M. OSMO$^1$, A. PALLAVICINI$^1$, \underline{F. PICCININI}$^1$}

\address{$^1$Dipartimento di Fisica Nucleare e Teorica, Universit\`a di Pavia \\
and INFN Sezione di Pavia, via A. Bassi 6, Pavia, Italy; \\
$^2$Dipartimento di Fisica, Universit\`a di Ferrara \\
and INFN Sezione di Ferrara, Ferrara, Italy}

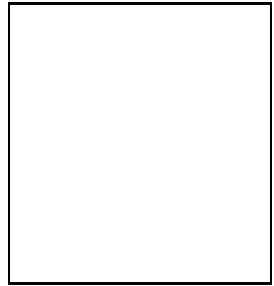
\maketitle\abstracts{The LEP data on four fermion processes are presently 
used to study the non abelian gauge couplings of the Standard Model. 
The present theoretical error for two classes of processes, single-$W$ 
production and radiative four fermion final states, is discussed according 
to the results of the four fermion working group of the LEP2 Monte Carlo 
Workshop held at CERN. }

\section{Introduction}

The center of mass (c.m.) energy reached at LEP2 allows to test 
directly the non abelian gauge couplings in $e^+ e^-$ collisions. 
The most copiously produced final states are four fermion final states, which, 
apart from being very important in the measurement of the $W$ properties 
such as mass and width~\cite{venturi}, contain information on the 
interactions among the 
gauge vector bosons.
In order to study 
the trilinear gauge couplings, also other processes are presently 
considered, such as single-$W$ production 
and 
$e^+ e^- \to \nu {\bar \nu} \gamma$. 
These two processes 
are very sensitive to 
the $W W \gamma$ vertex and are 
useful to disentangle the $W W Z$ from the $W W \gamma$ interactions 
present in the four fermion final states. At present also hypothetical 
self-interactions among the neutral vector bosons are looked for in the 
LEP experiments through the analysis of the neutral current four fermion 
final states and the radiative fermion pair production. 
At LEP there is also the opportunity to directly study, for the first time in 
$e^+ e^-$ collisions, the quartic gauge couplings, which are 
a window on the simmetry breaking mechanism. In particular, the quartic 
vertices involving at least one photon can be studied through the analysis of 
the radiative four fermion processes, the $\gamma\gamma + E\!\!\!\!/$~~and the
$f {\bar f} \gamma \gamma$ final states. 
In order to establish possible departures from the Standard Model (SM) 
predictions of the 
gauge boson self-interactions, it is extremely important to have sufficiently 
precise theoretical predictions to exploit the experimental precision. 
For this reason, within the {\it four fermion 
working group} of the LEP2 Monte Carlo Workshop~\cite{yr2k}, the $4f$, 
single-$W$ and $4f + \gamma$ final states have been studied with the aim of 
assessing the 
present theoretical accuracy of the SM predictions. In the present 
contribution some issues related to single-$W$ and radiative four fermion 
final states will be reviewed.

\section{Single-$W$}

Being the foreseen accuracy of final LEP2 data of the order of 
few per cent~\cite{yr2k}, all the theoretical effects on this scale need 
to be taken into account in the calculations. 
Due to the particular kinematical configuration with one electron/positron in 
the very forward region, the single-$W$ 
process is very challenging from the theoretical 
point of view. Actually it requires a massive treatment of the phase space 
and of the matrix element, due to strong cancellations in the region of the 
very small electron/positron scattering angles.

Several programs have been developed to treat this particular process, such as 
{\tt CompHEP}~\cite{comphep}, {\tt GRACE}~\cite{grace}, 
{\tt NEXTCALIBUR}~\cite{nextcalibur}, {\tt SWAP}~\cite{qedsc}, 
{\tt WPHACT}~\cite{wphact}, {\tt WTO}~\cite{wto}, each of them adopting an 
independent approach for the calculation 
of the matrix element and of the kinematics. Such a pletora of programs 
allowed to reach a very high technical precision, at the 0.1\% level, 
after careful 
tuned comparisons. 
In order to avoid integration singularities, it is mandatory to include 
the gauge boson width in the propagator. In general this 
introduces a violation of gauge invariance. 
This problem has been extensively studied~\cite{fl} 
and several options to address it have been explored. The most theoretically 
appealing procedure is the fermion loop scheme, which preserves $U(1)$ and 
$SU(2)$ Ward identities. Recently, this scheme has been generalized to the case 
of massive external fermions, both in its minimal version, which considers the 
imaginary parts of the fermion loops (IFL)~\cite{ifl}, and in its full 
realization with real and imaginary parts~\cite{efl}. In particular, a 
numerical investigation has been performed~\cite{ifl}, showing 
no significant difference between the IFL and the fixed width scheme, even in 
the region most sensible to $U(1)$ gauge invariance, thus justifying the use 
of the fixed width for simplicity reasons. 
The complete fermion loop scheme, contaning also the real part of the 
corrections, allows to evaluate the effects of the running couplings, 
which in the case of single-$W$ amount to about 5-7\%, depending on the 
channel and the adopted selection criteria~\cite{yr2k,efl}. 
However, being the leading dynamics given by the $W$ fusion diagram and by 
the $W$ bremsstrahlung diagram, with a $t$-channel photon, it is quite natural 
to ascribe the main contribution of the correction to the running of the 
electromagnetic coupling $\alpha_{QED}$. According to this, a simple 
prescription follows, which amounts to calculate the matrix element with the 
FW scheme, and rescale the differential cross section as 
$d\sigma/dt \rightarrow (\alpha^2(t)/\alpha^2_{G_F})\, d\sigma/dt$, 
where $\alpha_{G_F}$ is the electromagnetic coupling calculated according to 
the input parameter scheme, while $\alpha(t)$ is the coupling calculated at 
the tipical scale of the process. 
This very simple prescription have been proved to work very well for the 
semileptonic signature, with high invariant mass cut (which is a realistic 
event selection), while it differs of about 2\% for the fully leptonic sample, 
for the cuts which enhance the contribution of the perypheral 
diagrams~\cite{yr2k,efl}. This difference is however of the order of the 
intrinsic precision of the fermion loop scheme.

A further relevant issue is given by radiative corrections due to photon 
radiation. Given the particular kinematical configuration of the single-$W$ 
process with a charged particle lost in the beam pipe, the question naturally 
arises whether a Leading Log (LL) description is meaningful. 
Actually, by looking at 
the tree-level differential distribution of the virtual photon 
four-momentum transfer $t$, 
the largest part of the events are characterized by a ratio 
$t/m_e^2 \gg 1$. This 
allows to adopt the LL approximation, where, according to the factorization 
theorems, the QED corrected cross section of a generic process can be 
written as a convolution of the form
\begin{equation}
 \sigma = \prod_i \int  dx_i D(\Lambda_i^2,x_i) \; \sigma_0\, , 
\nonumber
\end{equation}
\noindent
where the index $i$ denotes a generic charged external line. 
The choice of the scales $\Lambda_i$ is not dictated by general arguments. 
A generally adopted attitude is given by the comparison of the 
$O(\alpha)$ expansion of the above convolution with a diagrammatic calculation 
which reproduces the correct LL contribution. 
Tipical simple examples are 
$e^+ e^- \to f {\bar f}$, with $f \neq e$ and Bhabha scattering, for which 
an exact $O(\alpha)$ perturbative calculation exist. 
In the case of single-$W$ production, the exact $O(\alpha)$ 
perturbative calculation is still missing, so a general strategy 
for the evaluation of the scales $\Lambda_i$ is needed. 
As a first step, a LL diagrammatic calculation, taking into 
account soft and collinear photon bremsstrahlung and its virtual counterpart, 
and, in the case of a calorimetric measurement 
of the energy of the final-state (FS) particles, also hard radiation 
collinear to the FS particles themselves, can be performed.
\noindent
Then the comparison between the result of such a calculation and the 
$O(\alpha)$ expansion of the SF QED corrected cross section allows to 
fix the scales $\Lambda_i$. 
If a calorimetric measurement of the energies of the FS particles is 
performed, only the IS legs need to be corrected by
the SF's. Furthermore, since the electron is scattered in the very
forward region, the interference between the electron line and the rest of the
process is very small. This allows a natural sharing of the logarithms 
between the two SF's associated to the colliding 
electron and positron, whose scales read~\cite{yr2k,qedsc} 

\begin{equation}
  \Lambda_-^2 = 4E^2 \frac{(1-c_-)^2}{\delta^2} \;,\quad
  \Lambda_+^2 = 2^{\frac{14}{9}}E^2
          \frac{\big((1-c_{\bar d})(1-c_u)^2\big)^\frac{2}{3}}
          {\big((1-c_{u{\bar d}})^2\delta^5\big)^\frac{2}{9}}, \nonumber
\end{equation}
where $\delta$ is the half-opening angle of the calorimetric resolution.
By using the above scales in the calculation of the total cross section leads 
to a QED correction of about 7-8\%, depending on the c.m. energy, 
lower (higher) of about 4\% with respect to a calculation which adoptes 
$s$ ($t$) as scale for both SF's. Similar quantitative conclusions have 
been reached independently by the {\tt GRACE} group~\cite{yr2k}. 

To summarize, the total theoretical error associated with single-$W$ 
production has been conservatively estimated within the four fermion 
working group~\cite{yr2k} to be $\pm 5\%$, even if some authors consider 
realistic a theoretical error of the order of 3\%.

\section{Radiative four fermion processes}

Also in the case of $4f + \gamma $ final states, several computational 
tools have been developed, such as  {\tt CompHEP}~\cite{comphep}, 
{\tt GRACE}~\cite{grace}, 
{\tt NEXTCALIBUR}~\cite{nextcalibur}, 
{\tt PHEGAS/HELAC}~\cite{phegashelac}, {\tt RacoonWW}~\cite{ddrw},
{\tt WRAP}~\cite{4fgpv}. By means of these programs a technical precision 
at the 0.1\% level has been reached in cross sections as well as 
distributions~\cite{yr2k}.
\noindent
The tuned comparisons have been performed by adopting the massless 
approximation for the outgoing fermions. However, fermionic mass terms can 
become important, due to the collinear ``singularities'' associated with 
photons emitted from the external legs, i.e. fermion mass effects are expected 
to be relevant for small angular separation cuts photon-charged fermions.
For instance, for the process 
$e^+ e^- \to \mu^- \bar{\nu}_{\mu} c \bar{s} \gamma $ at $\sqrt{s} = 200$~GeV 
and with a minimum angle of $5^\circ$ between photon and quarks, the relative 
difference between a massless calculation and a massive one is $1.9\%$ for 
$\theta_{min}^{\gamma - \mu} = 1^\circ$, and reaches $9.3\%$ for 
$\theta_{min}^{\gamma - \mu} = 0.1^\circ$.
Moreover, in the case of a final state muon, the separation cut can 
realistically be even $0^\circ$, and only 
a massive calculation can deal with this phase space region.

Due to the presence of an observed photon in the final state, 
the treatment of initial state radiation (ISR)
(which in the LEP2 energy range gives a reduction of the 
cross section of the order of 15\%) in terms of collinear SF's 
can become inadequate because affected by multiple counting
between the pre-emission photons, described by the SF's, and the observed one, 
described by the hard-scattering matrix element. 
A general strategy to deal with such a problem has been proposed for the 
$\nu {\bar \nu} \gamma$ final states, by using $p_\perp$ dependent 
SF's~\cite{nunugpv}. The same approach has been adopted in the code 
{\tt WRAP}~\cite{4fgpv}, to deal with the signature $4f + \gamma$. As 
a result of a preliminary investigation, i.e. by neglecting the contribution 
of final state radiation, the multiple counting contained in the predictions 
obtained with collinear SF's, affects the theoretical prediction at the 
5\% level with a minimum angle between charged fermions and photon of 
$10^\circ$ and $E_\gamma^{min} = 1$~GeV~\cite{yr2k,4fgpv}. 
Summarizing, the present theoretical error on radiative four fermion final 
states has 
been estimated~\cite{yr2k} to be at the 2.5\% level, due to the unknown 
missing non logarithmic terms of an exact $O(\alpha)$ calculation, 
to be compared with a final expected experimental error at the 5\% level, 
obtained by combining the data of the four LEP experiments.

\section{Conclusions}
At LEP there is the opportunity to directly measure the trilinear and 
quadrilinear gauge couplings, through the analysis of several processes.
In order to fully exploit the experimental precision, the theoretical 
predictions should be known with the highest possible accuracy. To this aim, 
at CERN it has been held in 1999 the four fermion working group of the 
LEP2 Monte-Carlo Workshop, where 
the present theoretical accuracy of the (SM) predictions 
has been discussed for several processes. In this contribution some 
theoretical aspects concerning single-$W$ production and radiative 
four fermion final states have been reviewed. In particular, as far as 
single-$W$ is concerned, the issue of gauge invariance, the effect of 
the running couplings and the proper scales in the SF's for the 
treatment of ISR have been discussed, summarizing the most recent work 
yielding a conservative estimate of the theoretical error of $\pm 5\%$. 
Regarding the $4f + \gamma $ final 
states, the effects of finite fermion masses and the problem of the 
correct treatment of ISR have been pointed out. The resulting theoretical 
error can be estimated to be of the order of 2.5\%.

\section*{Acknowledgments}
F.~Piccinini would like to thank the organizers, especially 
B.~Pietrzyk, for the kind invitation and the pleasant atmosphere 
of the Conference.

\end{document}